\documentclass{ws-procs9x6}

\usepackage{latexsym}                            
\usepackage{amsfonts}                            
\usepackage{amssymb}                             
\usepackage{epsfig}                              
\usepackage{amsmath}                              
\usepackage{wasysym}
\usepackage{slashed}

\renewcommand\draftnote{\hfill\hbox to \trimwidth{\footnotesize
\hfill\ \hfill}}%
\renewcommand\cropmarks{ }
\let\trimmarks\cropmarks

\begin{document}

\title{Polarized Structure Functions and the GDH Integral from Lattice
QCD\footnote{\uppercase{I}nvited Talk given at \uppercase{GDH}2004,
  \uppercase{J}une 
2-5, 2004, \uppercase{N}orfolk, \uppercase{U.S.A.}}}

\author{G. Schierholz}

\address{John von Neumann-Institut f\"ur Computing NIC,\\ Deutsches
  Elektronen-Synchrotron DESY,\\ 15738 Zeuthen, Germany\\and\\ Deutsches
  Elektronen-Synchrotron DESY,\\ 22603 Hamburg, Germany\\[1ex]
  E-mail: Gerrit.Schierholz@desy.de}  

\maketitle

\abstracts{
The Gerasimov-Drell-Hearn integral $I_{GDH}(Q^2)$, and its relation to
polarized nucleon structure functions, is discussed from the lattice
perspective. Of particular interest is the variation of $I_{GDH}(Q^2)$ with
$Q^2$, and what it may teach us about the origin and magnitude of higher-twist
contributions.} 

\vspace*{-10cm}
\noindent
DESY 04-198
\vspace*{9cm}

\section{Introduction}

The Gerasimov-Drell-Hearn (GDH) integral, which is written as
\begin{equation}
\begin{split}
I_{\rm GDH}(Q^2) &=\int_{\nu_0}^\infty \frac{d\nu}{\nu} \left[
  \sigma^{\stackrel{\rightarrow}{\Leftarrow}}(\nu,Q^2) -
  \sigma^{\stackrel{\rightarrow}{\Rightarrow}}(\nu,Q^2)\right]\\
&=\frac{8\pi^2 \alpha}{Q^2} \int_0^{x_0} dx\, \frac{1}{\sqrt{1+\gamma^2}}\,
  \tilde{A}_1 F_1 \\ &=
 \frac{16\pi^2 \alpha}{Q^2}\int_0^{x_0} dx\,
  \frac{g_1(x,Q^2) -\gamma^2
  g_2(x,Q^2)}{\sqrt{1+\gamma^2}} \,, 
\end{split}
\label{igdh}
\end{equation}
where $\nu_0=m_\pi+({m_\pi^2+Q^2})/{2 m_N}$, $x_0={Q^2}/{2 m_N \nu_0}$ and
$\gamma^2={4 m_N^2 x^2}/{Q^2}$,
connects the GDH sum rule at $Q^2=0$ to the Bjorken and Ellis-Jaffe
sum rules at large 
values of $Q^2$. The spin asymmetry $\tilde{A}_1$ is known over a large
kinematical region for 
proton, deuterium and helium targets, which allows to compute
$I_{\rm GDH}(Q^2)$ down to $Q^2 \approx 1$ GeV$^2$, separately for the proton
and the neutron. 

The GDH integral is of phenomenological interest for several reasons. It
involves both polarized structure functions of the nucleon, $g_1$ and $g_2$,
and thus tests the spin structure of the proton and the neutron. Furthermore,
the GDH 
integral provides a link between the nucleon state at high and at low
resolution, 
allowing us to study the transition from an assembly of quasi-free partons to 
strongly coupled quarks and gluons. In particular, we hope to learn about
the structure and magnitude of higher-twist contributions. This requires,
however, that
higher-twist contributions set in gradually and before $Q^2$ reaches
$\approx$~1~GeV$^2$, that is before the operator product expansion breaks
down.  
In order to match the predictions for $I_{\rm GDH}(0)$, the GDH sum rule, with
the Bjorken and 
Ellis-Jaffe sum rules, a strong variation of $I_{\rm GDH}(Q^2)$ with
increasing $Q^2$ is anticipated. 

Lattice QCD is in the position to address these questions. In this talk I
shall confront measurements of $I_{\rm GDH}(Q^2)$ with recent
lattice results.

\section{Polarized Structure Functions}

Let me recapitulate what we know about the nucleon's polarized structure
functions $g_1$ and $g_2$, which enter in (\ref{igdh}), first.

A direct theoretical calculation of structure functions is not possible. Using
the operator product expansion, we may relate moments of structure functions
in a twist or Taylor expansion in $1/Q^2$,
\begin{eqnarray}
2\int_0^1 dx\,x^n g_1(x,Q^2) &=& \frac{1}{2}\, e_{1,n}(Q^2/\mu^2,g(\mu^2))\,
a_n(\mu) +  O(1/Q^2)\,, \label{ope1}\\[0.5em]
2\int_0^1 dx\,x^n g_2(x,Q^2) &=& \frac{n}{2(n+1)}
\left[e_{2,n}(Q^2/\mu^2,g(\mu^2))\,d_n(\mu)\right. \label{ope} \\[0.5em]
&-& \left.e_{1,n}(Q^2/\mu^2,g(\mu^2))\,a_n(\mu)\right] + O(1/Q^2)\,, \nonumber
\end{eqnarray}
to certain matrix elements of local operators
\begin{equation}
\begin{split}
\langle \vec{p},\vec{s}|
{\mathcal O}^{5}_{ \{ \sigma\mu_1\cdots\mu_n \} }| \vec{p},\vec{s} \rangle
&= \frac{1}{n+1}a_n^q\, [ s_\sigma p_{\mu_1} \cdots
p_{\mu_n}
+ \cdots -\mbox{traces}]\,, \\[0.5em]
\langle \vec{p},\vec{s}| {\mathcal O}^{5}_{ [  \sigma \{ \mu_1 ] \cdots
\mu_n \} } | \vec{p},\vec{s} \rangle
&= \frac{1}{n+1}d_n^{\,q}\, [ (s_\sigma p_{\mu_1} -
s_{\mu_1}p_\sigma) p_{\mu_2}\cdots p_{\mu_n} \\
&\hspace*{3.85cm} + \cdots -\mbox{traces}]\,, 
\end{split}
\end{equation}
where
\begin{equation}
{\mathcal O}^{5}_{\sigma\mu_1\cdots\mu_n}
= \left(\frac{i}{2}\right)^n\bar{q}\gamma_{\sigma} \gamma_5
\mbox{\parbox[b]{0cm}{$D$}\raisebox{1.7ex}{$\leftrightarrow$}}_{\mu_1}
\cdots \mbox{\parbox[b]{0cm}{$D$}\raisebox{1.7ex}{$\leftrightarrow$}}_{\mu_n}
q -\mbox{traces}\,.
\end{equation}
In parton model language
\begin{equation}
\vspace*{0.25cm}
a_n^q=2 \int_0^1 d x\, x^{n} \large[\raisebox{-0.35cm}{$\underbrace{
\raisebox{0.35cm}{$q_\uparrow(x,\mu^2)
      -q_\downarrow(x,\mu^2)$}}$}\large]
=2\Delta^n q \,,
\end{equation}

\vspace*{-0.45cm}\hspace*{5cm} \raisebox{2ex}{$\Delta q(x,\mu^2)$}

\noindent
in particular $a_0^u=2\Delta u$, $a_0^d=2\Delta d$, while $d_n^q$ has twist
three and no parton model interpretation.

In the following I will restrict myself to nonsinglet and valence quark
distributions due to lack of space. These quantities show little difference
between quenched and full QCD calculations, so that I can further restrict
myself to quenched results. 

Let us first look at the structure function $g_1$. In Table~1 I compare the
lattice results for the lower moments of $\Delta q(x,Q^2)$~\cite{QCDSF} with
the corresponding phenomenological (experimental) numbers~\cite{BB}. The
quoted result for $\Delta u -\Delta d \equiv g_A$ has been taken from a
recent, `proper' extrapolation to the chiral limit~\cite{HPW},
shown in Fig.~1. By and large we find good agreement. 

Let us next look at the structure function $g_2$. This differs from $g_1$ by
twist-three contributions. From (\ref{ope}) we
readily see that $g_2$ fulfills the Burkhardt-Cottingham sum rule
\begin{equation}
\int_0^1 dx\, g_2(x,Q^2) = 0\,.
\end{equation} 

\begin{displaymath}
\begin{tabular}{c|l|l}
Moment & \multicolumn{1}{c|}{Lattice~\cite{QCDSF}} &
\multicolumn{1}{c}{Experiment~\cite{BB}} \\
\hline
$\Delta u_v$ & $\phantom{-}0.889(29)$ & $\phantom{-}0.926(71)$ \\
$\Delta d_v$ & $-0.236(27)$ & $-0.341(123)$ \\[1ex]
$\Delta^1 u_v$ & $\phantom{-}0.198(8)$ & $\phantom{-}0.163(14)$ \\
$\Delta^1 d_v$ & $-0.048(3)$ & $-0.047(21)$ \\[1ex]
$\Delta^2 u_v$ & $\phantom{-}0.041(9)$ & $\phantom{-}0.055(6)$ \\
$\Delta^2 d_v$ & $-0.028(3)$ & $-0.015(9)$ \\[1ex]
$\Delta u - \Delta d$ &
$\phantom{-}1.25(7)$ &
$\phantom{-}1.267(142)$
\\[1ex]
$\Delta^1 u - \Delta^1 d$ & $\phantom{-}0.246(9)$ & $\phantom{-}0.210(25)$
\\[1ex]
$\Delta^2 u - \Delta^2 d$ & $\phantom{-}0.069(9)$ & $\phantom{-}0.070(11)$\\
\hline
\end{tabular}
\end{displaymath}\\
{\footnotesize Table 1. Comparison of lattice and experimental values of the
  lower moments of 
  $\Delta q(x,Q^2)$, defined in (6), in the $\overline{MS}$ scheme at $Q^2=4$ GeV$^2$. The
  subscript $v$ refers to valence quarks.}

\begin{figure}[t]
\begin{center}
\epsfig{file=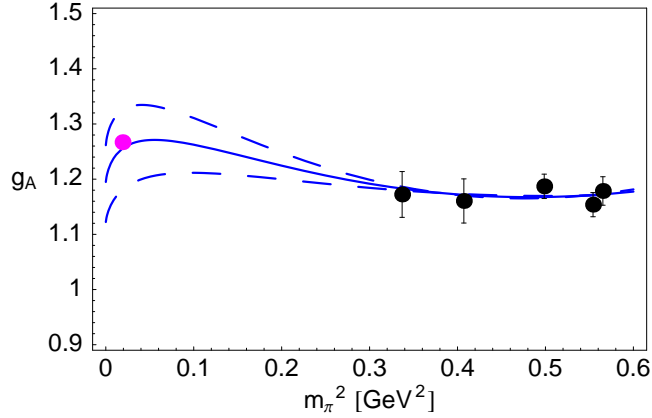,width=9cm,clip=}
\end{center}
\caption{The axial coupling $g_A$ as a function of $m_\pi^2$ together with a
fit from chiral perturbation theory.}
\end{figure}

\noindent
The first nontrivial moment of the twist-three contribution, $d_1^q$, can be
related to the tensor charge $\delta q$ of the nucleon~\cite{Paul},
\begin{equation}
\begin{split}
d_1^{\,q}\, (s_\mu p_\nu - s_\nu p_\mu) &= \langle \vec{p},\vec{s}|\bar{q}\!
\left(\gamma_\mu \gamma_5\,
\mbox{\parbox[b]{0cm}{$D$}\raisebox{1.7ex}{$\leftrightarrow$}}_{\!\nu}
- \mbox{\parbox[b]{0cm}{$D$}\raisebox{1.7ex}{$\leftrightarrow$}}_{\!\mu}
\gamma_\nu \gamma_5 \right)\!q | \vec{p},\vec{s} \rangle \\[0.75ex]
&= -\frac{i}{2}\, \langle \vec{p},\vec{s}|\bar{q}\!
\left(\sigma_{\mu\nu}\gamma_5 \,
\mbox{\parbox[b]{0cm}{$\slashed{D}$}\raisebox{1.7ex}{$\rightarrow$}}
+ \mbox{\parbox[b]{0cm}{$\slashed{D}$}\raisebox{1.7ex}{$\leftarrow$}}
\sigma_{\mu\nu} \gamma_5 \right)\!q | \vec{p},\vec{s} \rangle \\[0.75ex]
&= i\, m_q\, \langle \vec{p},\vec{s}|\bar{q}\sigma_{\mu\nu}\gamma_5 q|
  \vec{p},\vec{s} \rangle \\[0.75ex]
&= \frac{2 m_q}{m_N}\, \delta q\, (s_\mu p_\nu - s_\nu p_\mu)\,,
\end{split}
\end{equation}
where $m_q$ is the mass of the quark. Thus we have
\begin{equation}
d_1^q(Q^2) = \frac{2 m_q}{m_N} \delta q(Q^2)\,,
\end{equation}
which vanishes in the chiral limit ($m_q \rightarrow 0$). The second moment,
$d_2^q$, has been computed on the lattice~\cite{QCDSF2}. The result is shown
in Fig.~2, separately for the proton and the neutron, and found to be in good
agreement 
with experiment~\cite{dex}. From (\ref{ope1}) and (\ref{ope}) we obtain
\begin{equation}
\int_0^1 dx\, x^2 g_2(x,Q^2) + \frac{2}{3}\int_0^1 dx\, x^2 g_1(x,Q^2) =
\frac{1}{6} d_2\,.
\end{equation}
Given the fact that $d_2$ is small, and $d_1$ even vanishes in the chiral
limit, we derive that the Wandzura-Wilczek relation~\cite{WW} 
\begin{eqnarray}
\int_0^1 dx\, x^n g_2(x,Q^2) &=& -\frac{n}{n+1}\int_0^1 dx\, x^n
g_1(x,Q^2)\,,\\[0.5ex]
g_2(x,Q^2) &=& \int_{x}^1 \frac{dy}{y}\, g_1(y,Q^2) - g_1(x,Q^2)
\label{wand}
\end{eqnarray}
holds to better than $O(5\%)$, except perhaps at very large $x$, which we are
not 
interested in here. It is needless to say that $g_2$ in (\ref{wand}) satisfies
the Burkhardt-Cottingham sum rule as well. The structure function $g_1(x,Q^2)$
is 
obtained from the parton distributions $\Delta q(x,Q^2)$ by
\begin{equation}
g_1(x,Q^2) = \frac{1}{2} \sum_q e_q^2 \int_{x}^1 \frac{dy}{y}\,
e_1(y,Q^2)\,
\Delta q\left(\frac{x}{y},Q^2\right) 
\end{equation}
with
\begin{equation}
\int_0^1 dy \, y^n e_1(y,Q^2) = e_{1,n}(1,g(Q^2))\,.
\end{equation}

\begin{figure}[t]
\vspace*{-1.25cm}
\begin{center}
\epsfig{file=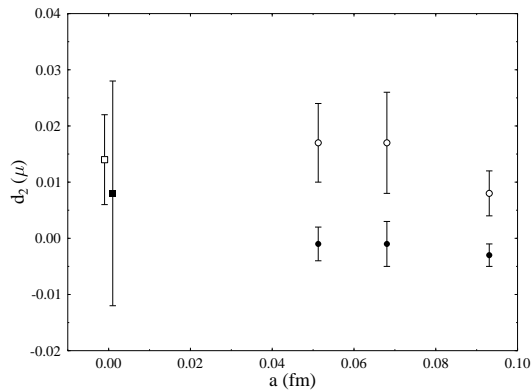,width=9cm,clip=}
\end{center}
\vspace*{-0.75cm}
\caption{The twist-three contribution $d_2(\mu)$ at $\mu^2=5$ GeV$^2$ for
proton ($\Circle$) and neutron ($\CIRCLE$) plotted against the lattice spacing
$a$, together with the experimental numbers for proton ($\Box$) and neutron
({\tiny $\blacksquare$}).}
\end{figure}

\section{Higher-Twist Contributions}

Contributions of twist four (and higher) have been studied on the
lattice, either through calculations of appropriate nucleon matrix
elements~\cite{ht}, or 
by evaluating the operator product expansion directly on the
lattice~\cite{htope}. Higher-twist contributions are 
generally found to be small. 
Four-fermion operators, for example, schematically drawn in
Fig.~3, which account for diquark effects in the nucleon, contribute 
\begin{equation}
\int_0^1 dx\, F_2^{(4)}(x,Q^2)\big|_{{\bf 27}, I=1} = - 0.0005(5)\,
\alpha_s(Q^2)\,
\frac{m_N^2}{Q^2} \,.
\end{equation}
The reason for quoting the flavor 27-plet structure function here, rather than
the nucleon (octet) one, is that the underlying four-fermion operator does
not mix with operators of lower dimension, which makes it a clean prediction.

\begin{figure}[t]
\vspace*{-1.35cm}
\begin{center}
\epsfig{file=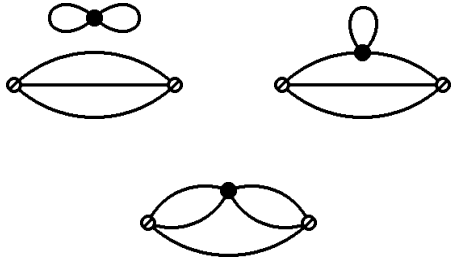,width=9cm,clip=}
\end{center}
\vspace*{-1.5cm}
\caption{Diagrams involving four-fermion operators.}
\end{figure}

\section{The GDH Integral}

We are ready now to address the GDH integral. Following our previous
discussion, we may assume that higher-twist contributions are small, and
that the nucleon's second polarized structure
function $g_2$ is well approximated by the Wandzura-Wilczek form
(\ref{wand}). Moreover, given the fact that the lattice predictions for
$a_n^q$, compiled in Table 1, are in good agreement with the phenomenological
numbers quoted, we may base our further discussion on the parameterization of
$g_1(x,Q^2)$ given in Ref.~2. 

As already mentioned, I will restrict myself to the nonsinglet GDH integral,
which corresponds to proton minus neutron target. In Fig.~4 I show recent
results from the Hermes 
collaboration~\cite{hermes}. I compare this result with the theoretical
predictions. The solid line represents the full integral, as given by the
bottom line of (\ref{igdh}),
including $g_1$ and $g_2$, while the dashed line represents the integral
\begin{equation}
 \frac{16\pi^2 \alpha}{Q^2}\int_0^1 dx\,
  g_1(x,Q^2) \equiv  \frac{16\pi^2 \alpha}{Q^2} \Gamma_1\,. 
\label{bj}
\end{equation}

\clearpage
\begin{figure}[!ht]
\vspace*{-0.25cm}
\begin{center}
\epsfig{file=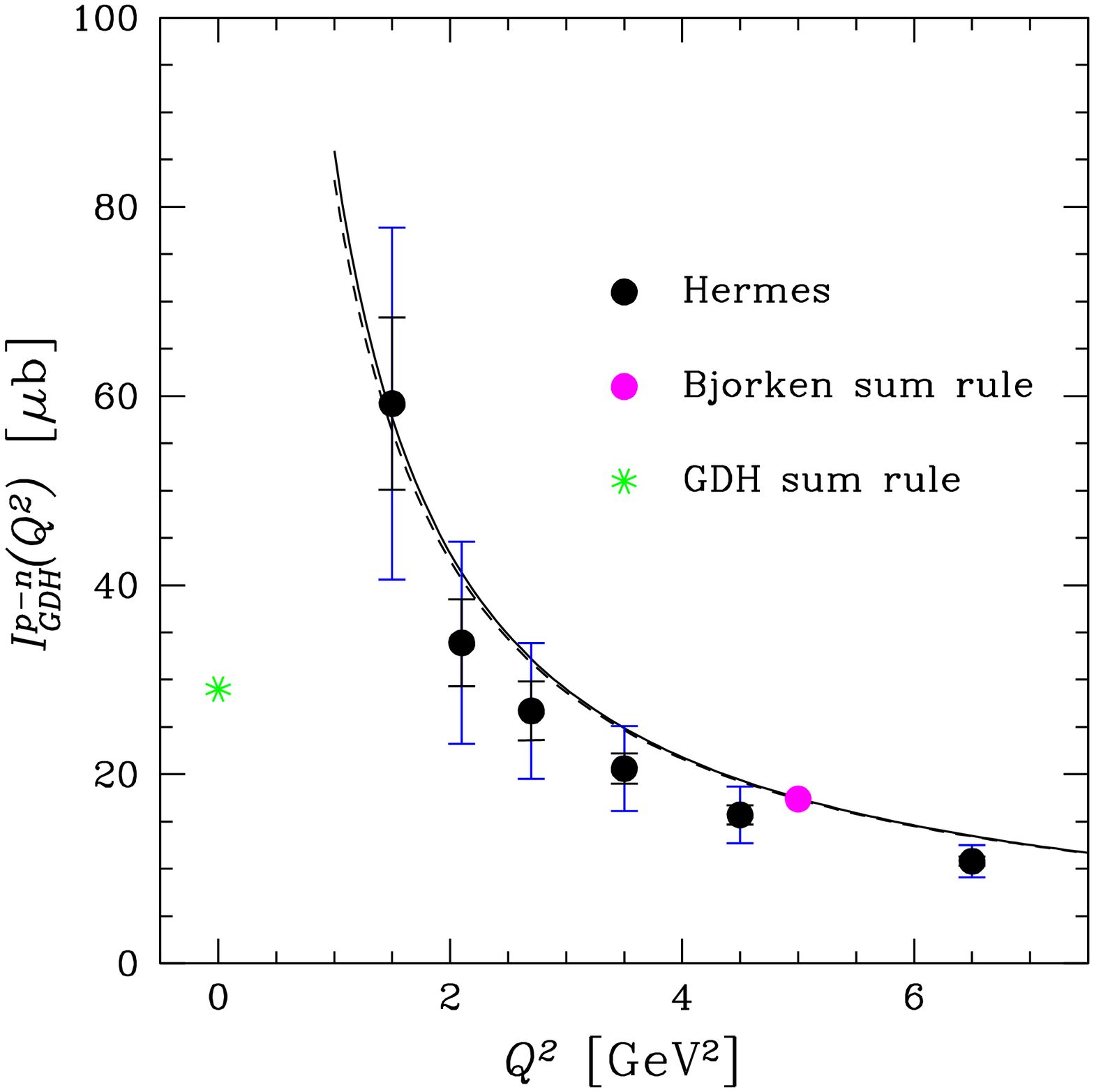,width=7.8cm,clip=}
\end{center}
\vspace*{-0.25cm}
\caption{The GDH integral: Experiment versus theory. The inner error bars are
  statistical, the outer ones statistical plus systematic.}
\vspace*{0.25cm}
\begin{center}
\epsfig{file=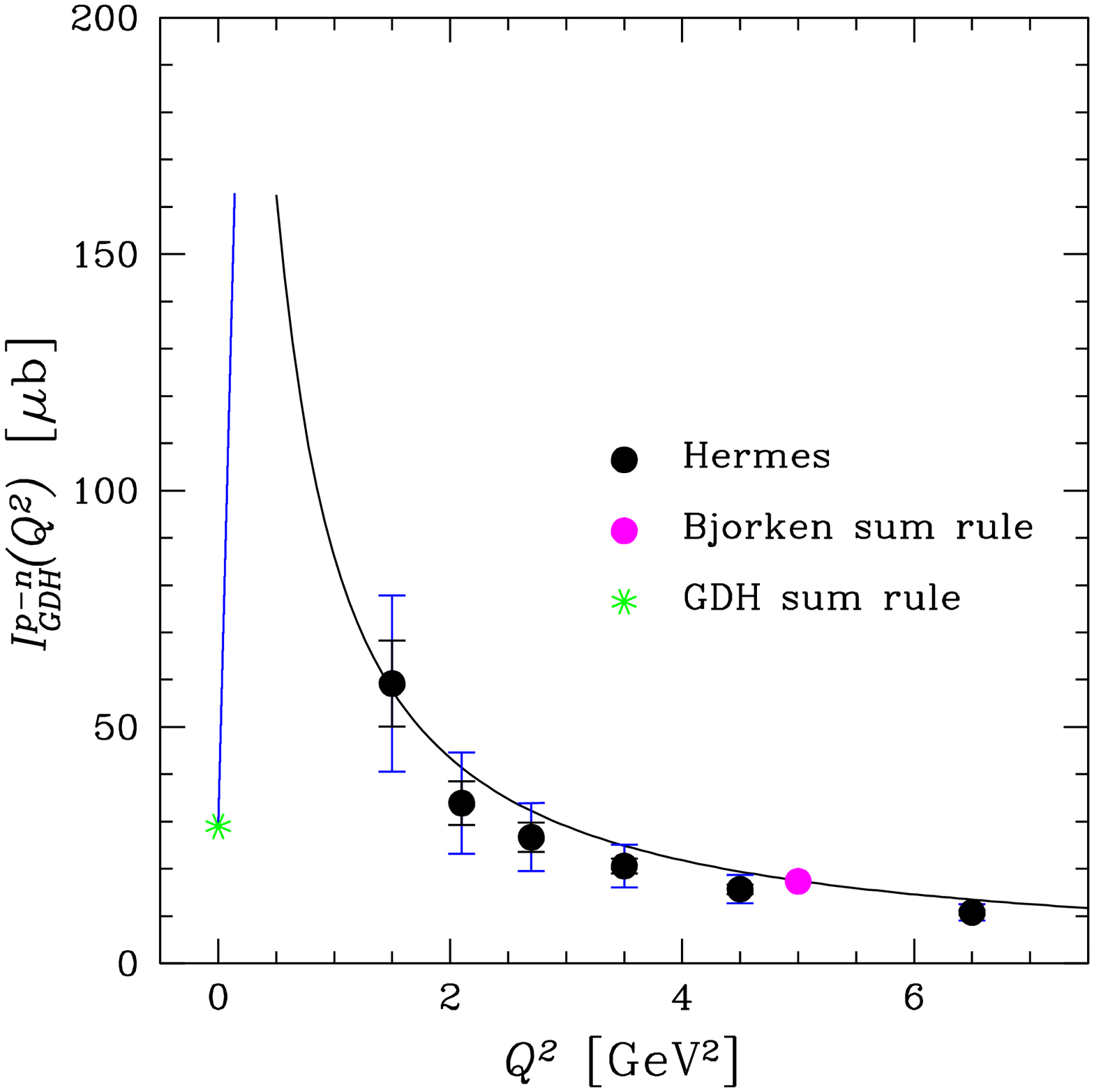,width=7.8cm,clip=}
\end{center}
\vspace*{-0.25cm}
\caption{The GDH integral: The connection to the GDH sum rule. The left curve
is the prediction of chiral perturbation theory.}
\end{figure}

\clearpage 
\noindent
Equation (\ref{bj}) corresponds to the limiting case $x_0 = 1$ and $\gamma =
0$. The 
dashed curve can hardly be distinguished from the solid curve over the whole
kinematical range, $1 \lesssim Q^2 \leq \infty$, which tells us that the
(nonsinglet) GDH integral is insensitive to $g_2$ and merely tests the Bjorken
sum rule
\begin{equation}
\begin{split}
\Gamma_1 = \int_0^1 dx\,  g_1(x,Q^2) &= \frac{1}{6} g_A
\left[1-\frac{\alpha_s(Q^2)}{\pi} 
-3.58\, \left(\frac{\alpha_s(Q^2)}{\pi}\right)^2\right. \\ &- 20.22\,
  \left.\left(\frac{\alpha_s(Q^2)}{\pi}\right)^3 + \cdots \right] \,.
\end{split}
\end{equation}

At larger values of $Q^2$ the data fall below the curve. This may be due to
the fact that the $x$ range covered shrinks to $0.2 \leq x \leq 0.8$ in the
highest-$Q^2$ bin. In order to draw quantitative conclusions, one certainly 
would need more precise data. So far we can only say that the experimental
data are consistent with leading-twist parton distributions all the way down
to $Q^2 \approx 1$ GeV$^2$, and with the Bjorken sum rule in particular.

In Fig.~5 I show the same figure on an enhanced scale, together with the
predictions of chiral perturbation theory~\cite{bhm} and the predicted value
of the GDH sume rule. Both curves appear to meet in a sharp peak 
at $Q^2 \approx 0.4$ GeV$^2$. This value lies much below the range of validity
of the operator product expansion. Likewise it lies beyond the applicability of
chiral perturbation theory, so that no firm statement about the transition
from large $Q^2$ to the resonance region can be made.  

\section{Conclusions}

To learn anything new, if possible at all, from the GDH integral, we
need better experimental data. At present the GDH integral does not teach us
anything 
quantitative about higher-twist contributions. Lattice calculations, 
on the other hand, indicate that the GDH integral is well represented by the
Bjorken sum rule down to $Q^2 \approx 1$ GeV$^2$. More precise
lattice data on moments of $g_1$ and $g_2$, including sea quark effects, will
become available soon~\cite{soon}.  
In order to match scaling and resonance region, it appears that one will have
to resort to model building.

\section*{Acknowledgement}

I like to thank Johannes Bl\"umlein, Helmut B\"ottcher, Meinulf G\"ockeler,
Roger Horsley, Dirk Pleiter and Paul Rakow for discussions and
collaboration. 



\begin{thebibliography}{99}
\bibitem{Paul} This derivation is due to Paul Rakow. Equation (9) is still to
  be verified numerically.
\end{thebibliography}
\end{document}